\documentclass[12pt]{iopart}
\usepackage{iopams}  

\usepackage{graphics}

\begin{document}
\title{Photon statistics characterization of a single photon source}
\author{R.~All\'eaume\dag, F.~Treussart\dag\footnote[3]{To whom correspondence should be addressed.}, J.-M.~Courty\ddag \ and J.-F.~Roch\dag}

\address{\dag
Laboratoire de Photonique Quantique et Mol\'eculaire\footnote[4]{Laboratoire du CNRS, UMR 8537, associ\'e ˆ l'\'Ecole Normale Sup\'erieure de Cachan}, ENS  Cachan, 61 avenue du pr\'esident Wilson, 94235 Cachan cedex, France}

\address{\ddag
Laboratoire Kastler Brossel\footnote[5]{UMR 8552, Unit\'e mixte de recherche de l'\'Ecole Normale Sup\'erieure, du CNRS, et de l'Universit\'e Pierre et Marie Curie}, UPMC case 74, 4 place Jussieu, 75252 Paris cedex 05, France}

\ead{francois.treussart@physique.ens-cachan.fr}

\begin{abstract}
In a recent experiment, we reported the time-domain intensity noise measurement of a single photon source relying on single molecule fluorescence control. In this article we present data processing, starting from photocount timestamps. The theoretical analytical expression of the time-dependent Mandel parameter $Q(T)$ of an intermittent single photon source is derived from ON$\leftrightarrow$OFF dynamics . Finally, source intensity noise analysis using the Mandel parameter is quantitatively compared to the usual approach relying on the time autocorrelation function, both methods yielding the same molecular dynamical parameters.
\end{abstract}

 % end of PACS codes

\submitto{\NJP}

\maketitle
\section{Introduction}
\label{intro}
Optical experiments at the level of single quantum emitters allow one to produce specific quantum states of light with photon statistics that deviate strongly from classical distributions~\cite{Kimble_77,Diedrich_PRL87}. Despite the experimental challenges of producing single photon states~\cite{Yamamoto_PRL94, Yamamoto_Nature99}, recent developments of quantum information theory have intensified interest in single photon sources. Realization of an efficient single-photon source (SPS)  is, for instance, a key-problem in quantum cryptography and could more generally be applied to quantum information processing \cite{Knill_NAT}.

Recent experiments  reported quantum key distribution (QKD) with polarisation encoding on  single photons~\cite{Alexios_PRL02,Waks_NAT02}. They revealed potential gain of such sources over systems relying on strongly attenuated laser pulses. However, in these experiments, the
actual performance of QKD is intrinsically linked to photon statistics of the single photon source~\cite{Lutkenhaus_PRA}.

Following the proposal of De Martini et al.~\cite{DeMartini96,Rosa_PRA00}, we
recently realized a SPS based upon pulsed excitation of a single molecule~\cite{FMT_PRL02}.  Among various experimental realizations of single photon sources~Ê\cite{Brunel_PRL99,Lounis00,Michler_Science_00,Alexios_EPJD,Santori_PRL_01,Moreau_APL_01,Yuan_02}, a molecular-based SPS presents
several advantages.  First, it can be driven at room temperature with a relatively simple setup which achieves global efficiency exceeding 5~$\%$ for single photon production and detection. Secondly, since the molecular fluorescence lifetime is a few nanoseconds, high repetition rate can potentially be used.  Finally, background-emitted photon intensity level is extremely low, for carefully
prepared samples.

At the single pulse timescale, the figure of merit of a SPS can be characterized by efficiency of delivering triggered photons to target and by the ratio of single photon to multiphoton pulses~\cite{Rosa_PRA00}. 
In reference~\cite{FMT_PRL02}, we extended this analysis to measurement of SPS noise properties over a wide integration timescale range. In the detection scheme, complete statistical information
is extracted from the ``photocount by photocount'' record.  We showed that measured photon statistics strongly deviates from Poisson law, therefore clearly exhibiting non-classical features.
 
In this article, we detail the steps of this work, from realization of a molecular-based SPS to extensive statistical analysis of detected photons.

\section{Single photon emission from a single molecule}
\label{SPS}
\subsection{Principle of the experiment}
\label{principle}
\begin{figure}
\centerline{\resizebox{0.75\columnwidth}{!}{
\includegraphics{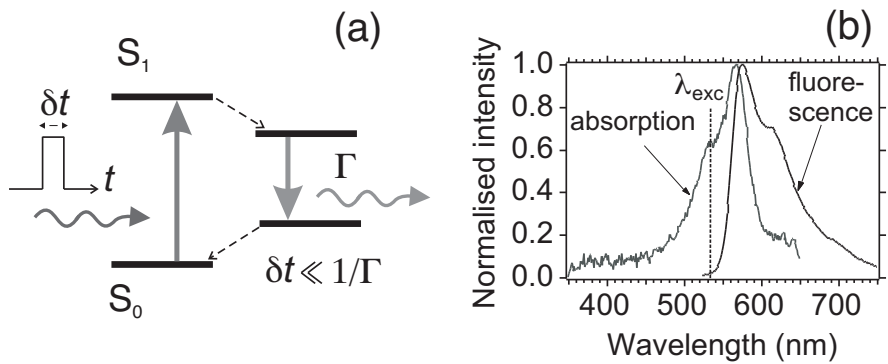}}}
\caption{(a) Single-photon generation by pulsed excitation of a single  4-level molecular system 
from the ground singlet state $S_{0}$ to a vibrationally excited  sublevel of the singlet state $S_{1}$.
Solid arrows corresponds to optical transitions, whereas dashed arrows depict non-radiative fast 
(ps) de-excitation. In order to emit a single photon per excitation pulse, the pulse duration $\delta t$ must be much shorter than the radiative lifetime $1/\Gamma$. 
(b) Absorption and fluorescence spectra of DiIC$_{18}$(3) dye embedded in a thin polymer film, measured respectively with absorption spectrometer and spectrofluorimeter with 514~nm excitation wavelength. Note that in the SPS experiment the 532~nm excitation wavelength is well separated from the dye's fluorescence emission which is centered at a wavelength around 570~nm FWHM.}
\label{mol_levels}
\end{figure}

As fluorescence light of a 4-level single emitter is antibunched for timescale on the order of
the excited state radiative lifetime~\cite{Alexios_EPJD,Kurtsiefer_00,Fleury_PRL00,FMT_OL01},
such systems can simply produce single photons on demand~\cite{DeMartini96,Rosa_PRA00,Lounis00}.
As summarized in figure~\ref{mol_levels}, the molecule is pumped into a vibrational excited state by a short excitation pulse. It then quickly decays to the first electronic excited state by a non-radiative process~\cite{Atkins_book}.  The emission of a single fluorescence photon then coincides with radiative de-excitation toward the ground state vibrational multiplicity, followed once again by fast non-radiative  decay (figure~\ref{mol_levels}). 
To emit more than one photon at a time, a molecule has to undergo a full excitation, emission and reexcitation cycle within the same excitation pulse. The probability of this occurrence is extremely small when the pulse duration is much shorter than the excited state lifetime~\cite{DeMartini96,Rosa_PRA00}.
Following the theoretical analysis of reference~\cite{Rosa_PRA00}, we
 chose a pulse duration of $\simeq150$~fs, which makes this probability less than
$5 \times 10^{-5}$. This value is negligible in comparison to the one associated to parasitic light, such as residual fluorescence from the molecular host matrix. To get one fluorescent photon per excitation pulse, the repetition period must also be much longer than the excited state lifetime so as to ensure relaxation into the ground state before application of the next excitation pulse.

%%%%%%%%%% fig.1 %%%%%%%%%%%%
%%%%%%%%%%%%%%%%%%%%%%%%
%%%%%%%%%%%% fig.2 %%%%%%%%%
\subsection{Experimental setup}
\begin{figure}
\centerline{\resizebox{0.75\columnwidth}{!}{
\includegraphics{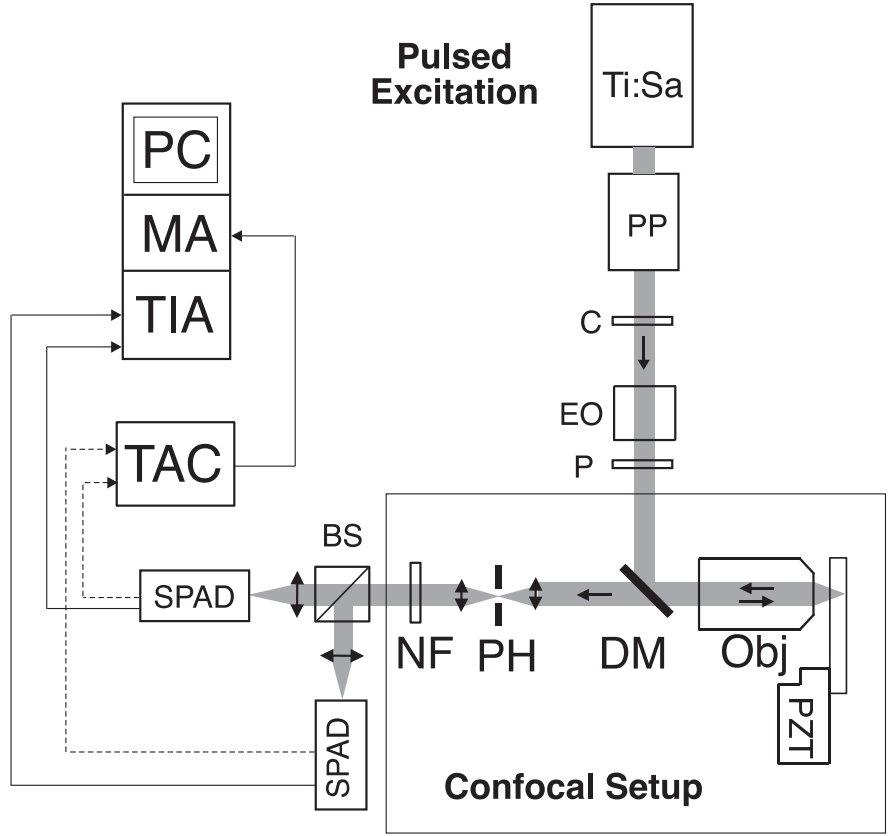}}}
\caption{Experimental setup for characterization of the triggered SPS. Single molecules are excited by a frequency-doubled femtosecond Ti:Sa laser at 532~nm. The laser is followed by PP: pulse picker; C: LiIO$_3$ nonlinear $\chi^{(2)}$ crystal; EO: ADP(NH$_4$PO$_4$) electro-optic cell; P: linear polarizer; PZT: piezoelectric translation stage; Obj: oil immersion microscope objective ($\times 60$, NA=1.4); DM: dichroic mirror;  PH: pinhole for confocal detection (30~$\mu$m diameter); NF: notch filter centered at 532~nm; BS: non-polarizing beamsplitter; SPAD: single photon 
silicium avalanche photodiode; TAC: time to amplitude converter; MA: multichannel analyser; TIA: time interval analyser (GuideTech, Model GT653) and PC: computer.}
\label{exp_setup}
\end{figure}
%%%%%%%%%%%%%%%%%%%%%%%%%

We use standard  confocal microscopy techniques to perform selective excitation and detection of single-molecule light emission at room temperature~\cite{FMT_OL01}. This setup  allows one to readily achieve two required features for observation of non-classical photon statistics, namely good collection efficiency of emitted photons and high rejection of optical background noise.

The laser source, used for fluorescence excitation, is a femtosecond tunable Titanium:Sapphire (Ti:Sa) laser, frequency doubled by single-pass propagation into a LiIO$_3$ nonlinear  crystal. The initial repetition rate of 82 MHz is divided by a pulse-picker with frequency set to 2.05 MHz (pulsed excitation repetition period $\tau_{\rm rep}=488~$ns) to avoid surpassing the maximum electronics counting rate.

The excitation light, centered at 532 nm, is reflected by a dichroic mirror into an inverted microscope. It is focused on the sample with an oil-immersion high numerical aperture objective, leading to a spot diameter of  $\simeq350$~nm FWHM.  The fluorescence light --redshifted with respect to the excitation-- is collected by the same objective, then transmitted through the dichroic mirror,  and finally focused inside a pinhole for the confocal configuration.  After recollimation, residual excitation light is removed by an holographic notch filter.

The samples used in our experiment consist of cyanine molecules DiIC$_{18}$(3).
This dye choice was motivated by its fluorescence efficiency and photostability, with an emission spectrum  well suited for detection using silicium avalanche photodiode (see figure~\ref{mol_levels} (b)). The dye molecules are embedded in a thin layer of PMMA deposited on a microscope coverplate by spin coating. The emitters are randomly distributed within the PMMA layer (thickness $\simeq30$~nm) at  an approximate concentration of one molecule per 10 $\mu \rm{m}^{2}$.

To ensure localization of a single emitter in the detection volume, we use a standard Hanbury Brown and Twiss setup \cite{Loudon}. It consists  of two single-photon-counting avalanche photodiodes (SPADs) placed on each side of a 50/50 nonpolarizing beamsplitter.  A Start--Stop technique with a time-to-amplitude converter allows us to build a coincidences histogram as a function of time delay between two consecutive photodetections on each side of the beamsplitter. Following the textbook experiment of P.~Grangier and A.~Aspect on quantum properties of single photon states  \cite{Grangier_EPL86}, the absence of coincidence at zero delay gives clear evidence of
single photon emission \cite{Brunel_PRL99}. 

We hence apply a simple three-steps procedure as explained in reference~\cite{FMT_PRL02}.  We first raster scan the sample at low energy per pulse ($\simeq0.5$ pJ) so as to map the fluorescence intensity and locate efficient emitters.  We then put the excitation beam on a given emitter and measure the autocorrelation function at low excitation energy.  We hence determine whether a single fluorophore or an aggregate of several molecules is excited.  Once such preliminary identification has been achieved, the single molecule is excited at much higher power so as to ensure saturated emission~\cite{FMT_PRL02}.
 
\subsection{Data acquisition}       
\label{data_acquisition}
Once a single emitter is located, we switch the detection procedure from the Start--Stop method to a complete recording of photon arrival times. The properly normalized interval function $c(\tau)$ measured by the Start--Stop technique corresponds to the intensity autocorrelation function $g^{(2)}(\tau)$ in the limit of short timescales and low detection efficiency~\cite{Reynaud_these_etat}.  However, to characterize more completely the statistical properties of the photon stream, one needs to test for correlations on timescales much longer than the excitation repetition period $\tau_{\rm rep}$.  In that case, the relationship between $g^{(2)}(\tau)$ and $c(\tau)$ becomes more complicated~\cite{Reynaud_these_etat,Fleury_PRL00}.  Instead of solely inferring $g^{(2)}(\tau)$ from $c(\tau)$ measurements, we have chosen to keep trace of the full range of dynamics by recording every photodetection time with a Time Interval Analyser (TIA) computer board. 
From this set of photocounts moments (that we call timestamps), detected photons statistics can then be directly analyzed over a wide range of timescales. Such a procedure avoids any mathematical bias in photon statistics analysis. 

\begin{figure}
\centerline{\resizebox{0.75\columnwidth}{!}{
\includegraphics{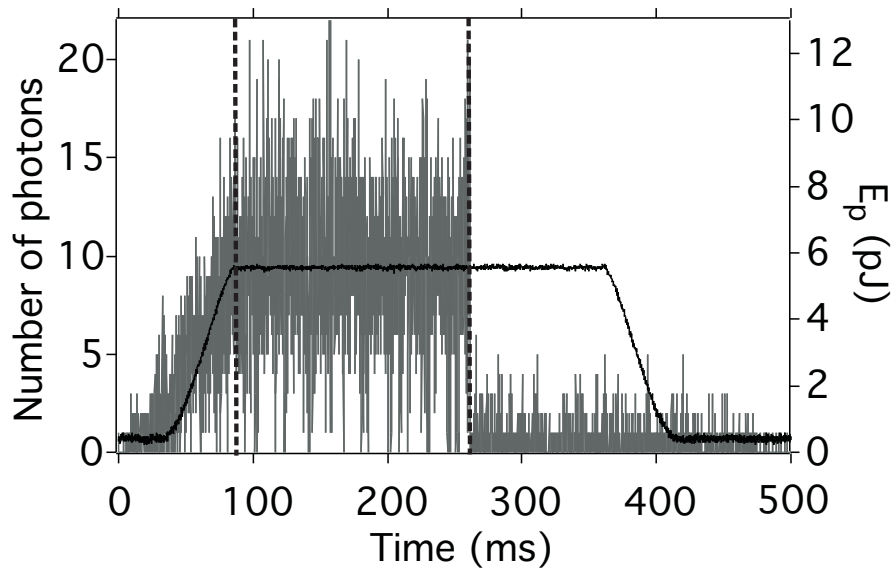}}}
\caption{Black solid line (right scale): laser pulse energy $E_{p}$ vs. time during single molecule excitation. Maximum laser pulse energy $E_{p}^{\rm max}$ of 5.6~pJ saturates the molecular transition.
Gray solid line (left scale): number of fluorescent photons detected during 50~$\mu$s integration duration. Photobleaching of the dye occurs 162~ms after excitation at the maximum energy per pulse begins, as delimited by two vertical dashed lines.}
\label{ramp+int}
\end{figure}

The total number of fluorescence photons that can be produced by a single molecule is limited at room temperature by its photostability \cite{Eggeling98}.  Under weak CW excitation, a molecule of DiIC$_{18}$(3) typically undergoes $10^{6}$ excitation cycles before irreversible photobleaching occurs \cite{Veerman_PRL99}. In our experiment, the excitation pulses energy is progressively ramped up to a maximum value of 5.6~pJ that ensures saturation of the $S_{0} \rightarrow S_{1}$ transition \cite{FMT_PRL02}.  This energy ramp, realized using an electro-optic modulator, consists of a 50~ms linear rise followed by a plateau lasting 300~ms and linear decrease (figure~\ref{ramp+int}).  We experimentally found that applying such a procedure substantially improves molecular photostability, compared to an abrupt excitation. 
We then select the timestamps of photocounts that occured during the plateau of the excitation. These events correspond to saturated emission of  our molecular SPS. Our analysis ``photocount by photocount'' then relies on determination of the number of detected photons in gated windows synchronized on Ti:Sa excitation pulses.

%%%%%%%%%%%% fig.3 %%%%%%%%%%%%%%%%%
%%%%%%%%%%%%%%%%%%%%%%%%%%%%%%%%

Data is first pre-processed over a discrete time grid. The excitation time base is
reconstructed from the timestamps ensemble $\lbrace t_i \rbrace$ and by applying a time filtering procedure described in \ref{analysistechnique}.  The gate duration is 30 ns, more than ten times the typical radiative lifetime of the molecule in PMMA layer.  All records outside the time gates are rejected, slightly improving the signal to background ratio.
Each timestamp $t_i$ is then attributed to a pulse $p_i$ in the time grid, and to each excitation pulse $p$, a number $n_p^{({\rm d})} = 0,1,2$ of detected photons is finally associated.  The probability distribution of the number of detected photons per pulse is deduced from $\{n_p^{({\rm d})} \}$, as summarized in table~\ref{StatResults}.

We next analyse the photon statistics of a data set extracted from SPS emission displayed in figure~\ref{ramp+int}.  We have selected photocounts recorded between the plateau beginning and molecular photobleaching clearly identified by a sudden drop in fluorescence emission.  During this time, the molecule was excited at constant maximum pumping energy, yielding 15332 photodetection events for 325313 excitation pulses.  The time filtering mentioned above is then applied keeping 15138 synchronous  photocounts .

\section{Single pulse photon statistics} 
\label{SPPS}
\begin{table}
\caption{\label{StatResults}Single-pulse statistics of a molecular single photon source, as obtained after numerical synchronization (see \ref{analysistechnique}). This data will be referred to as (S). The total number of excitation pulses in the sequence is 325313, leading to a total of  0,1 or 2 photon number events of 15138. The mean number of photons detected per pulse is $\langle n \rangle=0.04653$.  No events with $n^{\rm (d)}>2$ are observed due to deadtime in each detection channel.}
\begin{indented}
\item[]\begin{tabular}{@{}llll}
\br
$n^{\rm (d)}$(number of	& 0 			&	 1 		&	 2 			\\
 detected photons)  &&&\\
\mr
$n^{\rm (d)}$-photons 	&  310190		& 	15108	& 	15			\\
event number 		&&&\\
$n^{(d)}$-photon 		&  0.95351	& 	0.04644	 & 	$4.6\times10^{-5}$	 \\
event probability  	&&&\\
\br
\end{tabular}
\end{indented}
\end{table}

The single-pulse statistics presented in table~\ref{StatResults} are the direct outcome of photocounts acquisition. They correspond to the molecular emission displayed in figure~\ref{ramp+int}. While our SPS photon statistics appear to differ from a classical Poissonian distribution, the influence of our experimental setup on these measurements must be considered for accurate interpretation of these figures and for comparison with Poisson shotnoise reference.

\subsection{Influence of deadtime}
\label{deadtime}
In the following, we make a distinction between the distribution of the photons produced by the source, denoted by script notation ($\mathcal{P}$), and  photocount statistics, for which we conserve the usual notation ($P$).

Due to existence of a $\simeq280$-ns deadtime for each detection channel, a nonlinear relationship exists between detected photon statistics and source photon statistics.  Indeed, for a given excitation pulse, the number of detected photons in a 30-ns gated time window cannot exceed two if we use two avalanche photodiodes (APDs)  operating in the photon counting regime compulsory for our experiment.  

Denoting by $\mathcal{P}^{\mathrm{in}}(n)$ the photon number probability distribution of incoming light on the detection setup, the nonlinear tranformation relating this probability to the \emph{detected} photon probability $P(n=0,1,2)$  is simply computed for ``ideal APDs". By ``ideal APD", we mean that each photodiode clicks with 100 $\%$ efficiency immediately upon receiving a photon, but that no more than one click can occur in a given repetition period.  
In the approach developed here, we consider the ideal APD case for the following reasons:
\begin{itemize}
\item{limited quantum efficiency  of the APD (65~$\%$ in our experiment) is included in an overall linear loss coefficient along with other linear losses of the detection chain}
\item{deadtime is shorter than repetition period $\tau_{\mathrm{rep}}$ and much longer than pulse duration.}
\end{itemize}
For detection with a single ideal APD, the relationship between the photocount and incoming light statistics is
\begin{equation}
P(0) = \mathcal{P}^{\mathrm{in}}(0)\; {\rm and}\; P(1) = \sum_{n \geq 1}^{\infty} \,\mathcal{P}^{\mathrm{in}}(n) \,
\label{eq:1NL1}
\end{equation}

With our experimental detection scheme, random splitting of photons on two sides of 50/50 beamsplitter gives
\begin{eqnarray}
&P(0) =& \mathcal{P}^{\mathrm{in}}(0)  \label{eq:NL0} \\
&P(1) =& \sum_{n \geq 1}^{\infty} \,\mathcal{P}^{\mathrm{in}}(n) \,
\frac{1}{2^{n-1}}  \label{eq:NL1} \\
&P(2) =& \sum_{n \geq 2}^{\infty} \, \mathcal{P}^{\mathrm{in}}(n) \, \left(1-\frac{1}{%
2^{n-1}}\right)  \label{eq:NL2}
\end{eqnarray}

The relationship between $P(n)$ and photon statistics $\mathcal{P}(n)$  in SPS emission, comes from accounting for linear attenuation between SPS and detection.  We call $\eta$ the overall detection efficiency, which includes all linear propagation losses and  photodetector quantum efficiency.
$\mathcal{P}^{\mathrm{in}}$ is then related to $\mathcal{P}$ by the following binomial law
\begin{equation}
\mathcal{P}^{\rm in}(n) = \sum_{m=n}^{\infty} \left(^{m}_{n} \! \right) \eta^{n}
(1-\eta)^{m-n} \, {\mathcal{P}}(m)  \label{eq:Bernoulli}
\end{equation}

Combination of equations (\ref{eq:NL0}), (\ref{eq:NL1}), (\ref{eq:NL2}) and (\ref{eq:Bernoulli}) leads to a direct analytical relation between $\mathcal{P}(n)$ and $P(n=0,1,2)$.

Note that existence of such a saturation limit, due to detection deadtime, has no influence on photocount
statistics of a perfect SPS, for which $\mathcal{P}(n\ge 2)=0$, as long as excitation repetition period is longer than electronics deadtime.  On the contrary, for a ``real'' source with background light, the number of detected multi-photon pulses is systematically underestimated, leading to statistics artificially squeezed in comparison to shotnoise reference.

\subsection{Calibration with a coherent source}
\label{calibration}

\subsubsection{Coherent beam photocount statistics}
SPS performance can be directly evaluated by comparing single pulse photon statistics with those of a coherent source. This calibration takes into account the linear and nonlinear effects of our detection setup and permits accurate measurements of the multi-photon events probability reduction between single photon and Poissonian sources.

The photon number probability distribution for a coherent pulsed beam (C) is given by a Poisson law. According to equation~(\ref{eq:Bernoulli}), linear loss between the source and APDs change the mean photon number per pulse $\alpha$ to $\eta \alpha$ while the photon statistics remain Poissonian.  The expected photocount statistics can then be calculated by applying nonlinear transformations~(equations (\ref{eq:NL0}), (\ref{eq:NL1}) and (\ref{eq:NL2})) to a Poissonian distribution of parameter $\eta \alpha$
\begin{eqnarray}
&P_{\rm C}(0) &=e^{-\eta \alpha}  \label{eq:PC0} \\
&P_{\rm C}(1) &=2 e^{-\eta \alpha/2} (1 - e^{-\eta \alpha/2})  \label{eq:PC1} \\
&P_{\rm C}(2) &= (1- e^{-\eta \alpha/2})^{2},  \label{eq:PC2}
\end{eqnarray}
such distribution being termed as $P_{\rm C}(n)$ in table \ref{tablestat}.

\subsubsection{Experimental calibration}
A strongly attenuated pulsed laser beam is used as experimental reference to mimic a pulsed coherent source.  It is obtained by slightly detuning the Ti:Sa wavelength from the notch filter rejection band resulting in detection of residual pump light reflected from the sample.  The photocount statistics of this experimental reference (R) are then compared both with the experimental single photon source (S) and the calculated photocount distribution expected from a Poissonian source (C). To establish a valid comparison, calculated and experimental calibrations are determined for an --almost-- identical mean number of photons detected per pulse.

\begin{table}
\caption{\label{tablestat}Photocount probabilities $P(n)$ for SPS (S), reference experimental coherent source (R), and theoretical coherent source (C), for which photocount statistics are affected by detection. This table also displays the mean number $\langle n\rangle$ of detected photons per pulse.}
\begin{indented}
\item[]\begin{tabular}{@{}llll}
\br
& $n=1$ & $n=2$ & $\langle n\rangle$ \\ 
\mr
$P_{\mathrm{S}}(n)$ & 0.04644 & $\lineup{4.6\times10^{-5}}$ & 0.04653 \\
$P_{\mathrm{R}}(n)$ & 0.04520 & $50\times10^{-5}$ & 0.04620\\
$P_{\mathrm{C}}(n)$ & 0.04514 & $53\times10^{-5}$ & 0.04620 \\ 
\br
\end{tabular}
\end{indented}
\end{table}

Table~\ref{tablestat} shows that theoretical predictions for the coherent source are in good agreement with experimental calibrations, proving that our detection model accounts for all significant biases.  We can therefore confidently interpret the molecular SPS photon statistics we measure. 
For our SPS, the number of two-photons pulses is 10 times smaller than the corresponding probability for a Poissonian source. As mentioned earlier, residual multi-photon pulses mostly results from background fluorescence light triggered by Ti:Sa excitation. Indeed, the wavelength of this parasitic light lies within the molecule's fluorescence band and therefore cannot be filtered out. Careful optimization of substrate purity as well as of the chemicals purity used in sample fabrication can likely lower background fluorescence.
\subsection{Molecular SPS efficiency}
\label{efficiency}
Our molecular SPS emission can be modeled as the superposition of a perfect SPS and a coherent state of light.  In this model, all sources of linear loss (production~+~collection~+~detection), are gathered as an overall efficiency $\eta$.  For this perfect SPS (perf. SPS), the photon probability distribution of light inpinging on the APD is then given by
\begin{eqnarray}
&\mathcal{P}^{\rm in}_{\rm perf. SPS}(0) &= 1-\eta  \nonumber \\
& \mathcal{P}^{\rm in}_{\rm perf. SPS}(1) &= \eta  \label{sources_1}  \\
& \mathcal{P}^{\rm in}_{\rm perf. SPS}(n\ge2) &=0. \nonumber
\end{eqnarray}
Background (backgnd.) emission is modeled by a coherent state of light with a mean number 
$\eta\gamma$ of detected photons per pulse. The corresponding photon probability distribution is then
\begin{equation}
\mathcal{P}^{\rm in}_{\mathrm{backgnd.}}(n) = {e^{- \eta \gamma}(\eta \gamma)^{n}\over{n !}},\:{\rm for } \:n\ge 0.
\label{sources_2}
\end{equation}

Applying equations (\ref{eq:NL0}) to (\ref{eq:NL2}) to the (perf. SPS + backgnd.) probability distribution leads to the following analytical expressions for the real single photon source (S) \emph{photocounts} statistics~:
\begin{eqnarray}
&P_{\mathrm{S}}(0) &= e^{-\eta \gamma}\, (1-\eta)  \nonumber \\
& P_{\mathrm{S}}(1) &= 2\,( e^{-\eta \gamma/2} - e^{- \eta \gamma})+ \eta \, (2 e^{-\eta \gamma} - e^{- \eta \gamma/2})  \label{mol+fond} \\
& P_{\mathrm{S}}(2) &= (1-e^{-\eta \gamma/2})^{2} + \eta \,(e^{-\eta \gamma/2} - e^{- \eta \gamma}). \nonumber
\end{eqnarray}

Values for collection efficiency $\eta$ and signal-to-background ratio $1/ \gamma$ can be inferred from measured photocount statistics $P_{\rm S}$ (see table \ref{tablestat}).
Using equations (\ref{mol+fond}) for experimental values of $P_{\mathrm{S}}(1)$ and $P_{\mathrm{S}}(2)$, we finds $\eta\simeq0.04456$ and $\eta \gamma\simeq2.02\times 10^{-3}$. This leads to a signal-to-background ratio of 22, in good agreement with that measured by sample raster scan.

\subsection{Single-pulse Mandel parameter}
\label{Mandel}
From a statistical point of view there exists two main differences between experimental and ideal SPS: source overall efficiency lower than unity, and  finite ratio of single-photon to multi-photon pulses. Light produced by an ideal SPS consists in the periodic emission and detection of single photons with $100 \%$ efficiency, its intensity fluctuations being then perfectly squeezed. On the other hand, a real SPS yields less squeezing~\cite{Yamamoto_PRL94}. 

It is then meaningful to assess SPS performance by measuring its intensity noise on the
excitation repetition period $\tau_{\mathrm{rep}}$ timescale~\cite {FMT_PRL02}.
Such analysis requires evaluation of the single pulse Mandel parameter $Q$~\cite{Short_Mandel_PRL83}. This parameter characterizes deviation of photon statistics from Poissonian statistics for which $Q=0$. Subpoissonian (resp. superpoissonian) statistics correspond to $Q<0$ (resp. $Q>0$). For the distribution $\{n^{\rm (d)}_p\}$ of detected photon number, the Mandel parameter is defined by
\begin{equation}
Q \equiv \frac{\langle n^2 \rangle - {\langle n \rangle}^{2}}{\langle n\rangle} - 1  \label{defMandel}
\equiv {\langle (\Delta n)^2\rangle\over{\langle n \rangle}}-1,
\end{equation}
where $\langle n\rangle$ stands for the average value of $\{ {n^{\rm (d)}_p} \}$ calculated over the ensemble $\{p\}$ of excitation pulses. Note that an ideal SPS would yield $Q=-1$.
Moreover, for any statistical distribution, the effect of linear attenuation can be straightforwardly evaluated: after linear attenuation $\eta$, a Mandel parameter $Q_0$ would be changed in $\eta Q_0 $.  This means that every statistical distribution converges towards Poissonian statistics under attenuation.  This sensitivity to loss for measurements of non-zero Mandel parameters is similar to sensitivity  observed in squeezing experiments that measure reduced photocurrent noise spectra with respect to shotnoise reference.

For our molecular SPS, the Mandel parameter $Q$ of the \emph{photocount} statistics can be computed directly from single-pulse photocount probabilities
\begin{equation}
Q = \left[P(1) + 2 P(2) \right] \, \left\{ \frac{2 P(2) }{[ P(1)
+ 2 P(2) ]^{2} } - 1 \right\}.
\label{MandelExp}
\end{equation}
From table~\ref{tablestat}  data we infer a Mandel parameter $Q_{\rm S}= - 0.04455$ for the SPS.
This negative value for $Q$  confirms that our SPS indeed exhibits subpoissonnian statistics at the timescale $\tau_{\mathrm{rep}}$. Since very few multi-photon events are observed, the value of $Q_{\rm S}$ is almost only limited by the collection efficiency, which imposes a lower limit on $Q$: $Q_{\rm limit} = - \eta = -0.04456$.

Our measurement of $Q_{S}$ can then be compared to a Poissonian reference measurement.  Here again,  statistical bias introduced by APD deadtime must be taken into account. From equations~(\ref {eq:PC0})-(\ref{eq:PC2}) and (\ref{MandelExp}), we can derive the Mandel parameter of detected photons for a coherent source (C) of parameter $\alpha$.  Noticing that $\langle n\rangle_{\rm C}= 2 \, (1-e^{-\alpha/2} )$, we have
\begin{equation}
Q_{\rm C}= \langle n\rangle_{\rm C}\, 
\left[ \frac{2 P_{\rm C}(2) }{\langle n\rangle_{\rm C}^{2}}-1 \right] 
= - \frac{\langle n\rangle_{\rm C}}{2}.  
\label{MandelCoherent}
\end{equation}
As a consequence of photodetector deadtimes, a coherent source gives subpoissonian distribution of photodetection events. In our case, a coherent source with the same mean number $\langle n\rangle_{\rm C}=\langle n \rangle=0.04653$ of detected photons per pulse as the SPS, would then yield $Q_{\rm C}=-0.02327>Q_{\rm S}$.
Despite this detection bias, our direct measurement of the Mandel parameter, still yields a value for $Q_{\rm S}$ that clearly departs from that of Poissonian statistics. This \emph{measured} Mandel parameter is larger (in absolute value) than those measured in previous measurements by more than one order of magnitude~\cite{Short_Mandel_PRL83,Diedrich_PRL87,Fleury_PRL00}.

\section{Single photon source intensity fluctuations}
\label{intermittency}
Emission intermittency has been observed with most single photon sources realized so far~\cite{Lounis00,Alexios_EPJD,Santori_PRL_01}. This effect decreases source efficiency and contributes to additional source of noise. Better understanding of physical processes responsible for intermittency would likely lead to significant improvement of current SPS devices.

To characterize intermittency for our molecular SPS, we have investigated its influence on the photon statistics recorded with the time-resolved photon counting system. For a periodically trigerred SPS, this analysis is equivalent to study of source intensity noise over a wide range of timescales, which is usually done in the frequency domain for squeezing experiments, using a radio-frequency spectrum analyzer.

\subsection{Measuring intensity fluctuations : Time-varying Mandel parameter $Q(T)$}

The analysis performed on single-pulse photon statistics (see section~\ref{Mandel}) can be extended to multiple-pulses scale, allowing characterization of intensity fluctuations at any timescale greater than the pulse repetition period.
To do this, we analyze fluctuations of the total number $N\left( T\right)$ of photons detected during an integration time $T\equiv\mathcal{M\cdot }\tau _{\rm rep}$, which is a multiple of the repetition period. This analysis therefore corresponds to study of statistics of the photocounts number recorded during $\mathcal{M}$ successive pulses.

We then introduce the time-varying Mandel parameter $Q(T)$~\cite{Mandel_OL79}, defined similarly to single-pulse Mandel parameter. To perform statistical analysis, we extend the procedure used in section~\ref{SPPS}. More precisely, the complete data $\{n^{\rm(d)}_p\}_{p=1,\dots,\mathcal{N}}$ corresponding to photocounts recorded during $\mathcal{N}$ consecutive excitation pulses is split in successive samples, each lasting $T$.  We then obtain $\mathcal{N}_{\mathrm{sample}}=E\left[
\mathcal{N}/\mathcal{M}\right] $ samples.  
We call $N_{k}\left( T\right) $ the number of photocounts recorded during the $k^{\rm th}$ sample. We then have
\begin{equation}
N_{k}\left( T\right) \equiv\int_{kT}^{\left( k+1\right) T} I(t) dt =\sum_{p\,=k
\mathcal{M}}^{\left(k+1\right)\mathcal{M} -1}n_{p}.
\end{equation}
The statistical average over these samples of duration $T$ is denoted $\langle \;\rangle _{T}$, and we hence have
\begin{equation}
\left\langle N\right\rangle _{T}=\frac{1}{\mathcal{N}_{\mathrm{sample}}}
\sum_{k=0}^{\mathcal{N}_{\mathrm{sample}}-1}N_{k}\left( T\right).
\label{def_mean_N_T}
\end{equation}
Using this notations, the time-dependent Mandel parameter is given by
\begin{equation}
Q(T)\equiv \frac{\langle (\Delta N)^{2}\rangle _{T}}{\langle N\rangle _{T}}-1,
\label{QdeT}
\end{equation}
that allows direct comparison of SPS noise properties to those of Poissonian ligth beam.

\subsection{Intensity noise and intermittency in the molecular fluorescence: the ON-OFF model}
\label{model}
To analyze our experimental results and link them to physical parameters of molecular fluorescence, we use a simple analytical model of molecular intermittency, in which we assume that the SPS can be ON or OFF.  We call $p$ the ON to OFF transition rate and  $q$ the OFF to ON one. These rates correspond to lifetimes $\tau_{\rm on}=1/p$ and $\tau_{\rm off}=1/q$, respectively. 

\subsubsection{Physical interpretation of the ON-OFF model}
\begin{figure}
\kern 3mm
\centerline{\resizebox{0.75\columnwidth}{!}{
\includegraphics{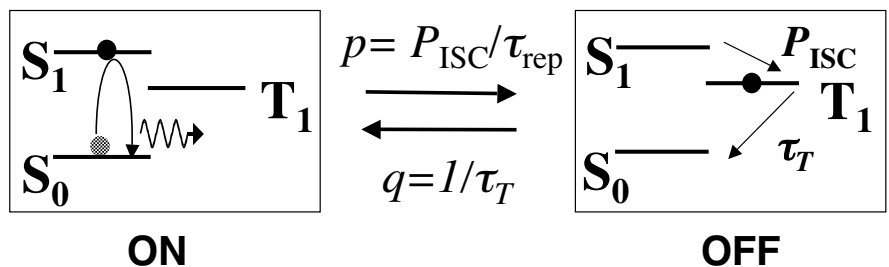}}}
\caption{Three states energy level structure and the corresponding ON and OFF states in the SPS intermittency model. In the ON state, the molecule undergoes fluorescent cycles between the ground $S_0$ and excited $S_1$ singlet states. In the OFF state, the molecule is trapped in the dark metastable  triplet $T_1$ state. Coupling from $S_1$ to $T_1$ occurs at each excitation pulse with the intersystem crossing probability $\mathcal{P}_{\rm ISC}$, yielding a transition ON$\rightarrow$OFF rate $p=\mathcal{P}_{\rm ISC}/\tau_{\rm rep}$. The reverse transition OFF$\rightarrow$ON occurs at rate $q=1/\tau_{\rm T}$, where $\tau_{\rm T}$ is the triplet state lifetime. }
\label{ON:OFF}
\end{figure}

For molecular SPS~\cite{FMT_PRL02} and other SPS's relying on fluorescence of a single emitter (e.g. single NV centers in diamond nanocrystals~\cite{Alexios_EPJD}), ON-OFF intermittency stems from the presence of a metastable non-fluorescent excited state in the energy level structure. 
Dynamics of ON-OFF behavior can then be computed from the three-level structure shown in figure~\ref{ON:OFF}
\begin{itemize}
\item ON $\rightarrow $ OFF transition corresponds to relaxation from the optical excited state $S_{1}$ to the triplet state $T_1$. For each excitation cycle, the probability $\mathcal{P}_{\rm ISC}$ of this intersystem crossing process is very small in the case of DiIC$_{18}$(3) molecule used in our experiment ($\mathcal{P}_{\rm ISC}\simeq10^{-4}$). 
Moreover, since singlet-triplet transitions occur exclusively from the excited state $S_1$, the excitation repetition period must be considered in defining the source ON state lifetime $\tau _{\rm on}$, which is then $\tau _{\rm on}=\tau _{\rm rep}/\mathcal{P}_{\rm ISC}$, assuming  saturated excitation regime.
\item  OFF $\rightarrow $ ON transition consists simply of non-radiative decay from triplet $T_1$ to ground $S_0$ state. Note that the triplet level is metastable since selections rules forbid direct optical transition to the ground state. The triplet state lifetime $\tau _{\rm T}=\tau_{\rm off}=1/q$ is therefore usually much longer than a typical fluorescent lifetime (in the case of DiIC$_{18}$(3), $\tau _{\rm T}\simeq200~\mu$s~\cite{Veerman_PRL99}).
\end{itemize} 
%%%%%%%%%%%%%%%% Fig. 4 %%%%%%%%%%%%%%%%%%
%%%%%%%%%%%%%%%%%%%%%%%%%%%%%%%%%%%%%%

\subsubsection{Dynamics of the ON-OFF system}
Under periodic pulsed excitation, the ON-OFF dynamics can be described using a discrete time model.  As transitions between ON and OFF states are random, we introduce a stochastic variable $r_{k}$  to account for the source state at instants $t_{k}=k \tau_{\rm rep}$. This parameter has value $r_{k}=1$ (resp. $r_{k}=0$) if the source is in the ON (resp. OFF) state at time $t_{k}$.

We then call $u_{k}$ the probability for the source to be in the ON state at time  $t_{k}$.  As the SPS emits photons exclusively from the ON state and never from the OFF state, $u_{k}$ also corresponds to the photoemission probability at time $t_{k}$.
We assume that lifetimes $\tau_{\rm on}$ and $\tau_{\rm off}$ of the ON-OFF states are much larger than the repetition rate $\tau_{\rm rep}$.  Then, the ON$\rightarrow$OFF transition probability is $p\tau_{\rm rep}$ and the OFF$\rightarrow$ON transition is $q\tau_{\rm rep}$.  It follows that the state of the emitter at pulse $k+1$ depends only on its state at pulse $k$. The recursion relation for the probability $u_{k+1}$ of the source to be ON at time $t_{k+1}$ is
\begin{equation}
u_{k+1}=\left( 1-p\tau _{\rm rep}\right) u_{k}+q\tau _{\rm rep}\,\left(
1-u_{k}\right),
\end{equation}
which leads to the general solution
\begin{equation}
u_{k}=\left( u_{0}-\frac{q}{p+q}\right) \,\left( 1-p\,\tau _{\rm rep}-q\,\tau_{\rm rep}\right) ^{k}+\frac{q}{p+q}.
\label{recur:2}
\end{equation}

Stationary probabilities for the molecule to be either ON or OFF are then
\begin{eqnarray}
&P_{\rm on}&={q\over{(p+q)}}\label{Pon} \\
&P_{\rm off}&=1-P_{\rm on}={p\over{(p+q)}}.\label{Poff}
\end{eqnarray}

\subsubsection{Source intensity and Mandel parameter vs. time}
\label{intensity_Q_T}
According to our model, light emitted by the source is a succession of single photon pulses emitted at time $t_{k}=k\tau _{\rm rep}$ with probability $u_{k}$, corresponding to intensity
\begin{equation}
I(t)=\sum_{k\,=\,-\infty }^{+\infty }\delta (t-k \tau _{\rm rep})\times r_{k},\; {\rm with}\; r_k=0\;{\rm or}\; 1.   
\label{int}
\end{equation}

The recursive relation~(\ref{recur:2}) for the ON-OFF model permits computation of statistical properties of source intensity $I(t)$.  In particular, we can derive the time-dependent Mandel
parameter from the variance of the number $N(T)$ of photons emitted by the intermittent source during  $T=\mathcal{M}. \tau_{\mathrm{rep}}$. Details of this calculation are given in \ref{CalcQdeT}.

Analytical expression of the Mandel parameter given by equation~(\ref{eq:QdeTcomplet}), can be simplified in the regime for which $\beta =(p+q)\tau _{\rm rep}\ll1$, leading to the following Mandel parameter expression for a ``perfect'' SPS with intermittency
\begin{equation}
Q_{\rm perf. SPS}(\mathcal{M}\tau _{\rm rep})=\frac{2p\times \tau _{\rm rep}}{\beta
^{2}} \left\{1- \frac{1}{\mathcal{M}\beta }\left[1-(1-\beta )^{\mathcal{M}}\right]\right\} -1.
\label{eq:QdeTsimplif}
\end{equation}

Experimental measurements of the Mandel parameter are also affected by overall efficiency $\eta $ smaller than unity  (see section \ref{efficiency}). Taking into account this limitation which is equivalent to linear loss, the Mandel parameter of the real source $Q_{\rm S}(T)$ is given by
\begin{equation}
Q_{\rm S}(T)=\eta \,Q_{\rm perf. SPS}(T).
\label{Q_linear_losses}
\end{equation}

\subsubsection{Experimental data analysis}
\label{expMandelT}
\begin{figure}
\centerline{\resizebox{0.75\columnwidth}{!}{
\includegraphics{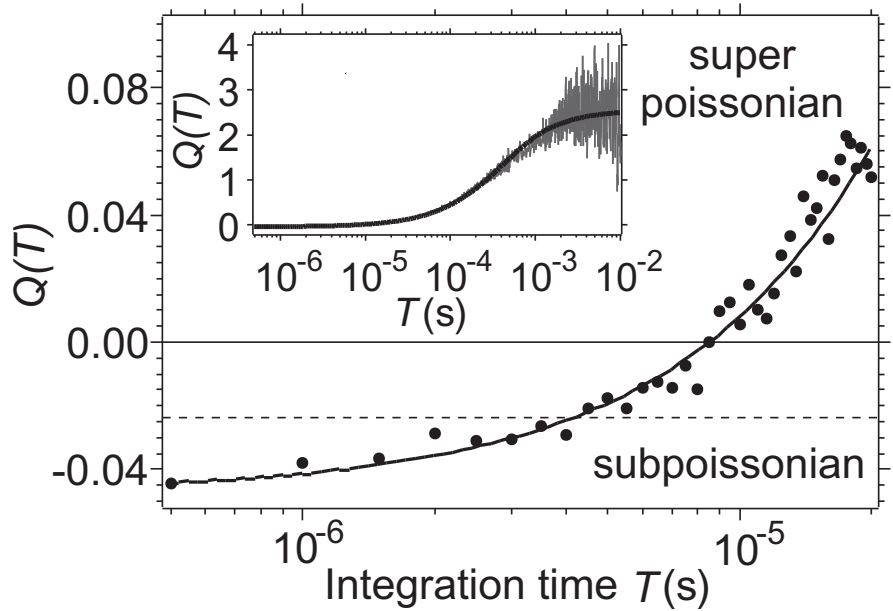}}}
\caption{Direct measurement of Mandel parameter $Q(T)$ over short integration time $T$. The dashed horizontal line shows $Q(T)$ for the equivalent coherent source (C), taking into account detection dead time. Inset shows $Q(T)$ for longer integration time. The solid curve is a fit  given by the model accounting for intermittency in SPS emission.}
\label{figmandel}
\end{figure}

As shown in figure~\ref{figmandel}, our experimental data are well-fitted by equations~(\ref{eq:QdeTsimplif}) and (\ref{Q_linear_losses}) over more than four orders of magnitude in time. Setting the measured efficiency to $\eta =0.04456$, the fit yields $p \tau _{\mathrm{rep}}=\mathcal{P}_{\rm ISC}=2.1\times 10^{-4}$ and $\tau _{\mathrm{T}}=250\,\mu$s, for the remaining two free parameters.
These values are in good agreement with values given in reference \cite{Veerman_PRL99}.

%%%%%%%%%%%%%% Fig. 5%%%%%%%%%%%%%%%
%%%%%%%%%%%%%%%%%%%%%%%%%%%%%%%%

Figure~\ref{figmandel} clearly shows that source photon statistics differ on short and long timescales. On timescales shorter than $\simeq8\tau _{\mathrm{rep}}$, the Mandel parameter of the source $Q(T)$ is smaller than $Q_{\rm C}$, the theoretical value of the Mandel parameter for poissonian light including the detection deadtime (horizontal dashed line on fig.\ref{figmandel}). On this short time scale, the SPS's photocount statistics are those of non-classical light.
On timescales larger than $\simeq10\mu s$, fluorescence intermittency  due to the triplet state, influences the photocount statistics by introducing excess of noise resulting in a positive value of the Mandel parameter.

The model developed here for a perfect intermittent SPS fits our experimental data with good accuracy. Indeed, apart from detection loss, other imperfections can be ignored or handle by the following:
\begin{itemize}
\item since the repetition period $\tau _{\mathrm{rep}}$ is much longer than the photodetection deadtime and since multi-photon events are extremely rare with our SPS, APD deadtime does not alter significantly the photocounts statistics. The detection can be considered effectively linear and equation~(\ref{Q_linear_losses}) remains valid in the presence of detection deadtimes.
\item high signal-to-background ratio means that background light does not contribute significantly to photocount statistics. It can therefore be neglected, as done implicitly in the model developped in this section. It can moreover be shown that addition of uncorrelated Poissonian background light of intensity $B$ to the perfect SPS signal $S$ is equivalent to loss. If we model the real source by the superposition of fluorescence background and light from a perfect SPS, then, introducing $\rho\equiv {S}/(S+B)$, the Mandel parameter of the real source is simply given by $Q_{\mathrm{S+B}}=\rho Q_{\mathrm{S}}$.
\end{itemize}

\subsubsection{SPS intensity autocorrelation function}
As a consistency check for our study, the time dependent Mandel parameter analysis can be compared  to a different approach using the intensity autocorrelation function $g^{(2)}$~\cite{Orrit_JCP93}, the measurement of which being at the heart of fluorescence correlation spectroscopy~\cite{Chu_91}.
%%%%%%%%%%%%%%% fig. 6 %%%%%%%%%%%%%%%%
\begin{figure}
\centerline{\resizebox{0.75\columnwidth}{!}{\includegraphics{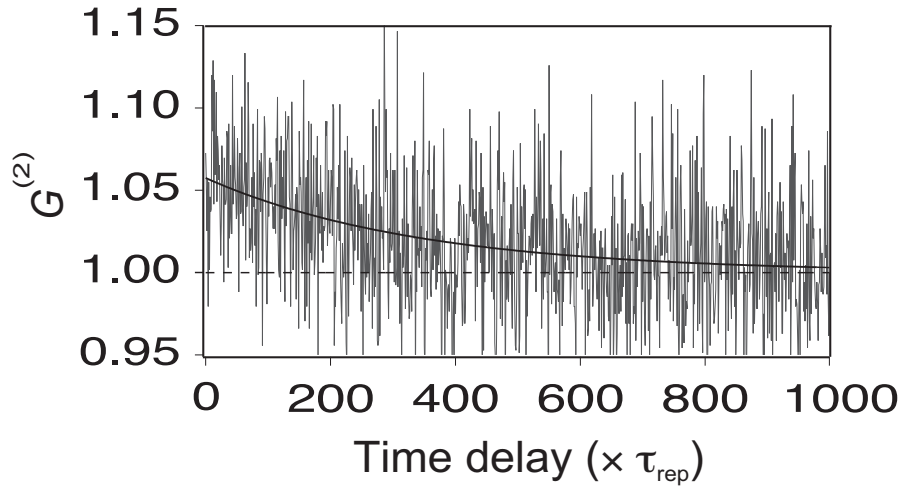}}}
\caption{Discrete-time photocount autocorrelation function $G^{(2)}$, computed from the set of data $\{n^{(d)}_p\}$ studied, in the range  of time delay $T = 1 \times \tau_{\rm rep}$ to $T= 1000 \times \tau_{\rm rep} $. The dashed horizontal line corresponds to shotnoise reference value.}
\label{fig:g2}
\end{figure}
%%%%%%%%%%%%%%%%%%%%%%%%%%%%%%%%%%

In terms of discrete time variables, a discrete time autocorrelation function for time delay $\Delta \times \tau_{\rm rep}$ is given by
\begin{equation}
G^{(2)}(\Delta)\equiv\frac{\langle n_{i} n_{i+\Delta}\rangle}{\langle n_{i}\rangle^{2}},
\label{eq:g2}
\end{equation}
where $n_{i}$ is the number of detected photons in the $i^{\rm th}$ excitation pulse, and $\Delta$ is an integer.

This discrete correlation function is directly related to the intensity autocorrelation function $g^{(2)}(\tau) \equiv\langle I(t)I(t+\tau)\rangle/\langle I(t) \rangle ^{2}$ usually measured with Start-Stop techniques~\cite{Brunel_PRL99}.  It can be shown that the normalized area of the $k^{\rm th}$ peak  (with the $0^{\rm th}$ reference peak corresponding to $\tau=0$) of $g^{(2)}(\tau)$ over a period $\tau_{\rm rep}$ is equal to $G^{(2)}(\Delta=k)$.

The ON-OFF model developed in section~\ref{intermittency} can be applied to calculate $G^{(2)}(\Delta)$
\begin{equation}
G^{(2)}(\Delta) =\frac{p}{q} e^{-(p+q) \Delta\times\tau_{\rm rep}},
\label{eq:g2Exp}
\end{equation}
which coincides with the formula given in reference~\cite{Santori_PRL_01}.

From our data $\{n^{(d)}_{p}\}$, we numerically compute $G^{(2)}(\Delta)$, varying $\Delta$ from 1 to 1000. Note that the latter value is chosen because blinking occurs in a timescale range of $\simeq1000\times \tau_{\rm rep}$. Results of this $G^{(2)}(\Delta)$ calculation are displayed in figure~\ref{fig:g2}. The experimental curve is fitted with equation (\ref{eq:g2Exp}), providing another way of measuring dynamical parameters of intermittent molecular SPS.
This fit yields $\mathcal{P}_{\rm ISC}= 1.6 \times 10^{-4}$ and $\tau_{\rm T}=180~\mu$s, which are in good agreements with the values obtained in section~\ref{expMandelT} using Mandel parameter analysis. 

Note that, on short time scale, the statistical noise is higher on $G^{(2)}$ than on $Q$. This is due to the fact that $G^{(2)}$ is computed over fewer but bigger statistical samples.

\section{Conclusion}
We have realized an efficient triggered single photon source relying on the temporal control of a single molecule fluorescence. After a comparison to Poissoninan coherent light pulses with the same mean number of photons per pulse, we have characterized intensity noise properties of this SPS in the time domain and photocounting regime.

From the record of every photocount timestamp, we calculate the second order correlation function $G^{(2)}$ or equivalently the time-dependant Mandel parameter $Q(T)$. Observed negative $Q(T)$ values signifie non-classical photocount statistics.

This time-domain analysis  is complementary to fluorescent correlation spectrocopy techniques for investigating photochemical properties at the single-emitter level. More specifically, we have modeled fluorescence intermittency by a two-state ON$\leftrightarrow$OFF dynamical process. By fitting a theoretical analytical expression of the Mandel parameter for an intermittent SPS, we obtained quantitative values for relevant molecular photodynamical parameters.
Such a direct time-domain statistical analysis could give insight into molecular properties such as conformational changes~\cite{Xie_ChemPhys02,Mukamel_JCP02}, resonant energy transfer \cite{Mabuchi_PRL02} or collective emission effects in multichromophoric systems~\cite{Basche_PRL03}.

With expected application to quantum cryptography, higher overall efficiency within a given emission spectral band should be reached so that single photon sources can exhibit advantages over attenuated laser pulses~\cite{Alexios_PRL02}. In recent experiments we coupled the fluorescence of a single emitter (a colored center in a diamond nanocrystal) to the single mode of a planar microcavity and observed a significant increase in spectral density of the emitted photons. These preliminary results are promising realization of an efficient single photon source well-suited for open-air quantum key distribution.
%\cite{Alleaume_NJP03}.

\ack{
The authors are grateful to  V.~Le Floc'h, L.T.~Xiao and C.~Grossman for their contributions at various point in the experiment. We also thank P.~Grangier and J.~Zyss for fruitful discussions, and Robin Smith for her valuable remarks on the manuscript. The experimental setup was built  thanks to the great technical assistance of A.~Clouqueur, J.-P.~Madrange and C.~Ollier. This work was supported by an ``ACI Jeune chercheur'' grant from Minist\`ere de la Recherche, and by a France T\'el\'ecom R\&D grant (``CTI T\'el\'ecom Quantique'').}

%%%%%%%%% APPENDICES %%%%%%%%%%%%
\appendix
%%%%%%%%%% Appendix A %%%%%%%%%%%%
\section{General analysis technique of a set of photocounts}
\label{analysistechnique} 
A set of data consists of a list of timestamps $\{t_i\}$ recorded by the Time Interval Analyser computer board. In this appendix we describe  the protocol developed to process raw data. This procedure allows us first to postsynchronize the timestamps on an excitation timebase and then to build the set $\{n^{(d)}_p\}$ of the number of detected photons for each excitation pulse $p$.

The pulsed excitation laser acts as a periodic trigger of emitted photons with repetition period $\tau _{ \mathrm{rep}}\simeq488~$ns.  An excitation laser pulse is emitted at time $t_{\mathrm{start} }+p\times \tau _{\mathrm{rep}},$ where the pulse is indexed by the integer $p$. The parameter $t_{\mathrm{start}}$ represents the pulse emission time taken as the first ($p=0$)  of the data.
For each set of data, $t_{\mathrm{start}}$ and $\tau _{\mathrm{rep}}$ must be determined because the repetition period of the Ti:Sa femtosecond laser can fluctuate  slightly between acquisitions. However, the laser repetition rate is stable over the typical acquisition duration (under one second), and $\tau _{\mathrm{rep}}$ is therefore constant for a given data record.

Single photon emission by the molecule occurs at each excitation pulse after a random time delay related to the molecule's excited state lifetime. Non-synchronous photocounts due to APD dark counts are rare, so almost all recorded photocounts are triggered by photons emitted by the molecule, with few by photons from residual fluorescence background. 
For these reasons, $t_{\mathrm{start}}$ and $\tau_{\mathrm{rep}}$ can then be determined directly from recorded data.

The $i^{\rm th}$ photocount timestamp $t_{i}$ can be expressed as
\begin{equation}
t_{i}=t_{\mathrm{start}}+ (p_{i} \times \tau_{\mathrm{rep}}) + \delta \tau_{i},
\label{ti}
\end{equation}
where integer $p_{i}  \in  \{ 1,\dots,\cal{N} \}$ indexes the laser pulse preceding detection of the $i^{\rm th}$ photon, and data to be analysed lasts $\cal{N}$ repetition periods; $\delta \tau _{i}$ is the time delay between excitation pulse  and photocount  timestamp ($0\leq \delta \tau _{i}<$ $\tau _{\mathrm{rep}}$ ).  
Given a set of timestamps $\{t_{i}\}$, relevant information can be equivalently represented by lists $\{p_{i}\}$ and $\{\delta \tau_{i}\}$, provided the value of $\tau _{ \mathrm{rep}}$ is known accurately enough.  Since the fluorescence lifetime of the molecule is much shorter than the laser repetition period, $\delta \tau _{i}\ll $ $\tau_{\mathrm{rep}}$, as long as photocount $i$ is not a dark count, which is rarely the case.

%%%%%%%%%%%% fig. A1%%%%%%%%%%%%%%%%%%%
\begin{figure}
\centerline{\resizebox{0.75\columnwidth}{!}{
\includegraphics{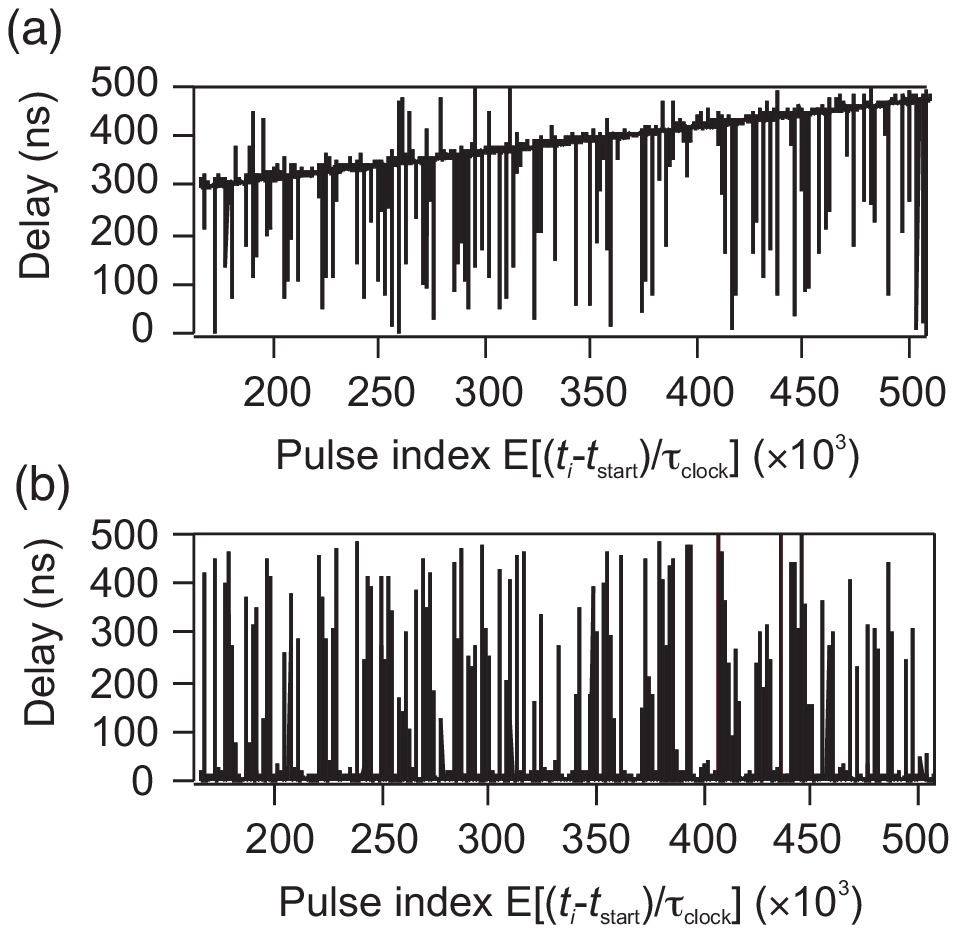}}}
\caption{Synchronization procedure of timestamps $\{t_i\}$, showing delay function $\mathrm{Delay}(t_i)$ vs. pulse index $E[(t_i-t_{\rm start})/\tau_{\rm clock}]$ for a given set of parameters $t_{\rm start}$ and $\tau_{\rm clock}$. (a) Case with a small linear drift of delay baseline, when $\tau_{\rm clock}$ is close to $\tau_{\rm rep}$ but $t_{\rm start}$ is incorrect. (b) Case when $\tau_{\rm clock}=\tau_{\rm rep}$ within relative precision of  $10^{-9}$ and $t_{\mathrm{start}}$ is properly estimated.}
\label{ecarts}
\end{figure}
%%%%%%%%%%%%%%%%%%%%%%%%%%%%%%%%%
As the laser period $\tau _{\mathrm{rep}}$ is not known precisely, we first attempt to synchronize the data on the excitation timebase considered as a clock of period $\tau_{\mathrm{ clock}}$ close to the expected laser period. We introduce a delay function parametrized with $t_{\mathrm{ start}}$ and $\tau _{\mathrm{clock}}$, that gives for each timestamp $t_i$ the time delay between this timestamp and the corresponding top of the clock\footnote{$E()$ stands for the integer part function.}
\begin{equation}
\mathrm{Delay}_{t_{\mathrm{start}},\tau_{\mathrm{clock}}}(t_{i})=
t_{i}-t_{\mathrm{start}}-E\left(\frac{t_{i}-t_{\mathrm{start}}}{\tau_{\mathrm{clock}}}\right)\times \tau_{\mathrm{clock}},
\end{equation}
where, if the clock period differs from the laser period, the drift of the time delay baseline with the pulse index $p_i$ is linear as ican be seen in figure~\ref{ecarts}(a)
\begin{equation}
\mathrm{Delay}_{t_{\mathrm{start}},\tau_{\mathrm{clock}}}(t_{i})
\simeq p_i (\tau _{\mathrm{rep}}-\tau _{\mathrm{clock}})
\simeq\frac{t_{i}}{\tau_{\mathrm{rep}}}\left( \tau _{\mathrm{rep}}-\tau _{\mathrm{clock}}\right).
\end{equation}
From the slope of the delay function baseline, we infer a new value for $\tau _{ \mathrm{clock}}$. Note that the first guess for $\tau_{\rm clock}$ is usually so far from the laser period that the delay function takes a saw-toothed shape, each jump corresponding to the delay reaching a multiple value of $\tau_{\rm clock}$.  As a consequence, only a linear fraction of the sample corresponding to a single saw tooth can be used at first.  
In further steps, estimation of $\tau _{\mathrm{rep}}$ improves and fewer jumps occur. Longer samples, corresponding to higher fit precision can then be processed. This procedure is repeated until the whole data set is used, leading to the situation of figure~\ref{ecarts}(b). It corresponds to the same fraction of data as in figure~\ref{ecarts}(a), for which $\tau_{\mathrm{rep}}$ is determined up to relative precision greater than $10^{-9}$.

Once $\tau_{\mathrm{rep}}$ and $t_{\mathrm{start}}$ values are known, calculation of the lists $\{p_{i}\}$ and $\{\delta \tau_{i}\}$ is straightforward using
\begin{equation}
p_{i} = E\left({t_i-t_{\rm start}\over\tau_{\mathrm{rep}}}\right)\; {\rm and}\;
\delta\tau_i = \mathrm{Delay}_{t_{\rm start},\tau_{\mathrm{rep}}}(t_i).
\end{equation}

A time filtering procedure is then used to eliminate all photocounts with time delay much longer than the molecule excited state lifetime. To implement this filter, we use a time window of duration $\Delta T_{\mathrm{window}}$. From the set $\{p_{i}\}$, we calculate the number $n_{i}$ of photons detected by the two photodiodes in the time interval $[p_{i}\tau_{\mathrm{rep}},\,p_{i}\tau _{\mathrm{rep} }+\Delta T_{\mathrm{window}}]$.  The time window duration  $\Delta T_{\mathrm{window}}$
must be shorter than the laser period and much longer than the molecular excited state lifetime $1/\Gamma$ so that the probability of discarding a ``real'' photodetection event is negligible.  The chosen time window duration $\Delta T_{\mathrm{window}}=30~$ns meets these two conditions, considering $1/\Gamma\simeq2.5~$ns for DiIC$_{18}$(3)  dye.  
Note that choice of a time window significantly shorter than the laser period has the advantage of filtering out our data from the majority of non-synchronous background photocounts, such as APD dark counts.

The processed data, now expressed as the table $\{n_{i},p_{i}\}$, and shortened to $\{n^{\rm (d)}_p\}$ in the body of this article, allows us to characterize the statistics of our source on timescales from $ \tau _{\mathrm{rep}}\simeq500$ ns to milliseconds.

%%%%%%% Appendix B %%%%%%%%%%
\section{Statistical characterisation of an intermittent SPS}
\label{CalcQdeT}
In this Appendix, we derive a general analytical expression of the ``perfect'' intermittent SPS Mandel parameter $Q(T)$ defined by equation~(\ref{QdeT}) using the ON-OFF model introduced in section~\ref{model}. We also retrieve the approximate expression~(\ref{eq:QdeTsimplif}) for $Q(T)$. We assume that source emission has reached its steady state at time $t=0$ .

The total number $N\left( T\right) $ of photocounts recorded  during an integration time $T=\mathcal{M\cdot }\tau _{\rm rep}$ corresponding to $\mathcal{M}$ consecutive excitation pulses is related to the stochastic photocount variable $r_k$ associated to $k^{\rm th}$ excitation pulse (see section~\ref{intensity_Q_T}) by
\begin{equation}
N\left( T\right) =\sum_{k\,=0}^{\mathcal{M}-1}r_{k},
\end{equation}
Calulating $Q(T)$ is equivalent to evaluating the variance ${\langle N^{2}\rangle }_{T}-{\langle N\rangle
_{T}}^{2}$, where mean values $\langle\;\rangle_T$ are defined by equation~(\ref{def_mean_N_T}). Both $\langle N\rangle_T$ and  $\langle N^2\rangle_T$ should then be evaluated. The mean value of $N(T)$ is given by:
\begin{equation}
\left\langle N\right\rangle _{T}=\sum_{k\,=0}^{\mathcal{M}-1}\left\langle
r_{k}\right\rangle =\mathcal{M}\frac{q}{p+q},
\label{mean_N}
\end{equation}
where the steady state expression of $\langle r_k\rangle=q/(p+q)$ comes from the definition of $r_k$ and the recursive law (\ref{recur:2}). Similarly, $\langle N^2\rangle_T$ follows
\begin{equation}
\langle N^{2}\rangle _{T}=\sum_{k=0}^{\mathcal{M}-1}\sum_{k\,^{\prime
}=0}^{ \mathcal{M}-1}\left\langle r_{k}\,r_{k^{\prime }}\right\rangle.
\end{equation}
Index change $\ell=|k-k'|$ in the previous equation then yields
\begin{eqnarray}
\langle N^{2}\rangle _{T} &=&\sum_{k\,=0}^{\mathcal{M}-1}\left\langle
r_{k}^{2}\right\rangle +2\sum_{k\,=0}^{\mathcal{M}-1}\sum_{\ell\,=1}^{
\mathcal{M}-k}C\left( \ell \right) \nonumber \\
&=&\sum_{k\,=0}^{\mathcal{M}-1}\left\langle r_{k}\right\rangle
+2\sum_{\ell \,=1}^{\mathcal{M}-1}\left( \mathcal{M}-\ell \right) C\left( \ell \right),\label{mean_N2}
\end{eqnarray}
where we have introduced the discrete time \emph{correlation function} $C(\ell)$ defined as
\begin{equation}
C(\ell) =\langle r_{k}r_{k+\ell}\rangle -\langle r_{k}\rangle \langle r_{k+\ell}\rangle.
\end{equation}

Calculation of $Q(T)$ now relies on evaluation of $C(\ell)$.  We recall that, in the model of section~\ref{model}, source dynamic is described by stochastic process $u_k$, the probability for the molecular system to be in the ON state. The general expression of $u_{k+\ell}$ follows from the recursive law~(\ref{recur:2})
\begin{equation}
u_{k+\ell}=\left(u_{k}-\frac{q}{p+q}\right)\,(1-p\,\tau _{\rm rep}-q\,\tau _{\rm rep})^{\ell}+\frac{q}{p+q}  \label{eq:evolution2}
\end{equation}

The stochastic variable $r_{k}$ equals 1 if the state is ON with probability $u_{k}$, and 0 if the state is OFF with probability $1-u_k$.
The product $r_{k}r_{k+\ell}$ is then equal to 1 only if both $r_{k}$ and $r_{k+\ell}$ are simultaneously equal to unity and otherwise equal to zero. It can be summarized as
\begin{eqnarray}
&\langle r_k r_{k+\ell} \rangle =P(r_k=1) P(r_{k+\ell}=1|r_k=1) \\
&\langle r_k \rangle \langle r_{k+\ell}\rangle=P(r_k=1) P(r_{k+\ell}=1).
\end{eqnarray}
Note that $P(r_k=1) $ is the probability that $r_k=1$, and $P(r_{k+\ell}=1|r_k=1)$ is the conditional probability for $r_{k+\ell}=1$ when $r_k=1$.  To fulfill this later condition $r_k=1$, one needs to have $u_k=1$. Moreover, by definition, the steady state probability is $P( r_{k}=1)=q/(p+q)$, so that we have
\begin{equation}
P(r_{k+\ell}=1|r_k=1) =\frac{p}{p+q}(1-p\tau_{\rm rep}-q\tau_{\rm rep})^{\ell}+\frac{q}{p+q},
\end{equation}
and, as a consequence,
\begin{equation}
C(\ell) =\frac{pq}{( p+q)^2}(1-p\tau_{\rm rep}-q\tau _{\rm rep})^\ell.
\end{equation}
This value $C(\ell)$ is then introduced in equation~(\ref{mean_N2}), and the expression for the variance follows from (\ref{mean_N2}) and (\ref{mean_N}). 
\begin{equation}
\langle N^{2}\rangle _{T}-\langle N\rangle
_{T}^{2}=\frac{pq}{(p+q)^{2}} \left[ \mathcal{M}\frac{1+\alpha
}{1-\alpha }-2\alpha \frac{1-\alpha ^{ \mathcal{M}}}{(1-\alpha
)^{2}}\right],
\end{equation}
where  $\alpha \equiv\left( 1-p\,\tau _{\rm rep}-q\,\tau _{\rm rep}\right).$

The general analytical expression of the ``perfect'' intermittent SPS Mandel parameter is finally deduced by
\begin{eqnarray}
&&Q_{\rm perf.SPS}(\mathcal{M} \tau _{\rm rep})={{\langle N^{2}\rangle_{T}-\langle N\rangle _{T}^{2}}\over{\langle N\rangle _{T}}}-1 \\
&=&\frac{p}{p+q}\left(\frac{2-\beta}{\beta}-\frac{2(1-\beta)}{\mathcal{M}} \cdot \frac{1-(1-\beta )^{\mathcal{M}}}{\beta^2}\right) -1,
\label{eq:QdeTcomplet}
\end{eqnarray}
where $\beta \equiv \left( p+q\right)\,\tau _{\rm rep}.$
In the limit of $\beta\ll1$, which is the case for molecular system dynamics considered in the body of this article, we retrieve expression (\ref{eq:QdeTsimplif}).

%%%%%%%%%
%\bibliography{pmu_romain}

\end{document}